\documentclass[11pt]{article}
\usepackage{moriond,epsfig}

\def\be{\begin{equation}}
\def\ee{\end{equation}}
\def\bea{\begin{eqnarray}}
\def\eea{\end{eqnarray}}

\newlength{\dslashwidth}

\newcommand{\bq}{\begin{equation}}
\newcommand{\eq}{\end{equation}}
\newcommand{\ba}{\begin{array}}
\newcommand{\ea}{\end{array}}
\newcommand{\bqa}{\begin{eqnarray}}
\newcommand{\eqa}{\end{eqnarray}}

\newcommand{\lnf}{{\ifmmode \Lambda^{(N_f)} \else $\Lambda^{(N_f)}$\fi}}
\newcommand{\ms}{{\ifmmode \overline{MS} \else $\overline{MS}$\fi}}
\newcommand{\dr}{{\ifmmode \overline{DR} \else $\overline{DR}$\fi}}
\newcommand{\lms}{{\ifmmode \Lambda^{(5)}_{\overline{MS}} \else $\Lambda^{(5)}_{\overline{MS}}$\fi}}
\newcommand{\lam}{{\ifmmode \Lambda \else $\Lambda$\fi}}
\newcommand{\gev}{{\ifmmode {\rm GeV} \else ${\rm GeV}$\fi}}
\newcommand{\gevc}{{\ifmmode {\rm GeV/c^2} \else ${\rm GeV/c^2}$\fi}}
\newcommand{\tev}{{\ifmmode {\rm TeV} \else ${\rm TeV}$\fi}}
\newcommand{\tevc}{{\ifmmode {\rm TeV/c^2} \else ${\rm TeV/c^2}$\fi}}
\newcommand{\lp}{{\ifmmode L^+  \else $L^+$\fi}}
\newcommand{\lm}{{\ifmmode L^-  \else $L^-$\fi}}
\newcommand{\mlp}{{\ifmmode M(L^-) \else $M(L^-)$\fi}}
\newcommand{\mlz}{{\ifmmode M(L^0) \else $M(L^0)$\fi}}
\newcommand{\lz}{{\ifmmode L^0 \else $L^0$\fi}}
\newcommand{\ev}{{\ifmmode GeV/c^2 \else $GeV/c^2$\fi}}
\newcommand{\tri}{{\ifmmode \triangleup \else $\triangleup$\fi}}
\newcommand{\unl}{{\ifmmode U_{lL^0} \else $U_{lL^0}$\fi}}\newcommand{\gL}{{\ifmmode g_L \else $g_{L}$\fi}}
\newcommand{\gR}{{\ifmmode g_R  \else $g_{R}$\fi}}
\newcommand{\gumu}{{\ifmmode \gamma^{\mu} \else $\gamma^{\mu}$\fi}}
\newcommand{\gunu}{{\ifmmode \gamma^{\nu} \else $\gamma^{\nu}$\fi}}
\newcommand{\gdmu}{{\ifmmode \gamma_{\mu} \else $\gamma_{\mu}$\fi}}
\newcommand{\gdnu}{{\ifmmode \gamma_{\nu} \else $\gamma_{\nu}$\fi}}
\newcommand{\stw}{{\ifmmode\sin^2\theta_W \else $\sin^{2}\theta_{W}$ \fi}}
\newcommand{\sws}{{\ifmmode \;\sin^2\theta_W  \else $\;\sin^{2}\theta_{W}$ \fi}}
\newcommand{\cws}{{\ifmmode \;\cos^2\theta_W  \else $\;\cos^{2}\theta_{W}$ \fi}}
\newcommand{\sw}{{\ifmmode \;\sin\theta_W  \else $\sin\theta_{W}$ \fi}}
\newcommand{\cw}{{\ifmmode \;\cos\theta_W  \else $\;\cos\theta_{W}$ \fi}}
\newcommand{\tw}{{\ifmmode \;\tan\theta_W  \else $\;\tan\theta_{W}$ \fi}}
\newcommand{\qq}{{\ifmmode q\overline{q} \else $q\overline{q}$\fi}}
\newcommand{\lR}{{\ifmmode l_R \else $l_R$\fi}}
\newcommand{\lL}{{\ifmmode l_L \else $l_L$\fi}}
\newcommand{\nt}{{\ifmmode \nu_{\tau} \else $\nu_{\tau}$\fi}}
\newcommand{\nuR}{{\ifmmode \nu_R  \else $\nu_R$\fi}}
\newcommand{\nuL}{{\ifmmode \nu_L  \else $\nu_L$\fi}}
\newcommand{\qR}{{\ifmmode g_R  \else $q_R$\fi}}
\newcommand{\qL}{{\ifmmode q_L  \else $q_L$\fi}}
\newcommand{\qRp}{{\ifmmode q_R'  \else $q_{R}$'\fi}}
\newcommand{\qLp}{{\ifmmode q_L'  \else $q_{L}$'\fi}}
\newcommand{\est}{{\ifmmode e^{\bf \ast} \else $e^{\bf \ast}$\fi}}
\newcommand{\lst}{{\ifmmode l^{\bf \ast} \else $l^{\bf \ast}$\fi}}
\newcommand{\must}{{\ifmmode \mu^{\bf \ast} \else $\mu^{\bf \ast}$\fi}}
\newcommand{\taust}{{\ifmmode \tau^{\bf \ast} \else $\tau^{\bf \ast}$ \fi}}
\newcommand{\pperp}{{\ifmmode p_t  \else $p_t$\fi}}
\newcommand{\et}{{\ifmmode E_t  \else $E_t$\fi}}
\newcommand{\xt}{{\ifmmode x_t  \else $x_t$\fi}}
\newcommand{\smumu}{{\ifmmode \sigma_{\mu\mu}  \else $\sigma_{\mu\mu}$ \fi}}
\newcommand{\eg}{{\ifmmode e\gamma  \else $e\gamma$\fi}}
\newcommand{\epem}{{\ifmmode e^+e^-  \else $e^+e^-$\fi}}
\newcommand{\lplm}{{\ifmmode L^+L^-  \else $L^+L^-$\fi}}
\newcommand{\pp}{{\ifmmode p\overline p  \else $p\overline p$\fi}}
\newcommand{\llz}{{\ifmmode L^0\overline{L}^0 \else $L^0\overline{L}^0$\fi}}
\newcommand{\epemt}{{\ifmmode e^+e^- \to  \else $e^+e^- \to$\fi}}
\newcommand{\eb}{{\ifmmode E_{beam}  \else $E_{beam}$\fi}}
\newcommand{\ip}{{\ifmmode pb^{-1}  \else $pb^{-1}$\fi}}
\newcommand{\upm}{{\ifmmode ^{\pm}  \else $^{\pm}$\fi}}
\newcommand{\de}{{\ifmmode ^{\circ}  \else $^{\circ}$ \fi}}
\newcommand{\appr}{{\ifmmode \sim \else $\sim$ \fi}}
\newcommand{\corresp}{{\ifmmode \stackrel{\wedge}{=} \else $\stackrel{\wedge}{=}$ \fi}}
\newcommand{\sqrts}{{\ifmmode \sqrt{s} \else $\sqrt{s}$\fi}}
\newcommand{\zz}{{\ifmmode Z^0  \else $Z^0$\fi}}
\newcommand{\mz}{{\ifmmode M_{Z}  \else $M_{Z}$\fi}}
\newcommand{\mzs}{{\ifmmode M_{Z}^2  \else $M_{Z}^2$\fi}}
\newcommand{\mw}{{\ifmmode M_{W}  \else $M_{W}$\fi}}
\newcommand{\mws}{{\ifmmode M_{W}^2  \else $M_{W}^2$\fi}}
\newcommand{\mh}{{\ifmmode M_{Higgs}  \else $M_{Higgs}$\fi}}
\newcommand{\gt}{{\ifmmode \Gamma_{tot} \else $\Gamma_{tot}$\fi}}
\newcommand{\msusy}{{\ifmmode M_{SUSY}  \else $M_{SUSY}$\fi}}
\newcommand{\msusys}{{\ifmmode M_{SUSY}^2  \else $M_{SUSY}^2$\fi}}
\newcommand{\su}{{\ifmmode SU(3)_C\otimes\- SU(2)_L\otimes\- U(1)_Y \else $SU(3)_C\otimes SU(2)_L\otimes U(1)_Y$\fi}}
\newcommand{\suthree}{{\ifmmode SU(3)_C  \else $SU(3)_C$\fi}}
\newcommand{\sutwo}{{\ifmmode  SU(2)_L\otimes U(1)_Y \else $SU(2)_L\otimes U(1)_Y$\fi}}
\newcommand{\taup} {{\ifmmode \tau_{proton} \else $\tau_{proton}$\fi}}
\newcommand{\as}{{\ifmmode \alpha_{s}  \else $\alpha_{s}$\fi}}
\newcommand{\mgut}{{\ifmmode M_{GUT}  \else $M_{GUT}$\fi}}
\newcommand{\mguts}{{\ifmmode M_{GUT}^2  \else $M_{GUT}^2$\fi}}
\newcommand{\mze} {{\ifmmode m_0        \else $m_0$\fi}}
\newcommand{\mha}{{\ifmmode m_{1/2}    \else $m_{1/2}$\fi}}
\newcommand{\mb} {{\ifmmode m_{b}    \else $m_{b}$\fi}}
\newcommand{\mt} {{\ifmmode m_{t}    \else $m_{t}$\fi}}
\newcommand{\mts} {{\ifmmode m_{t}^2    \else $m_{t}^2$\fi}}

\newcommand{\mtau}{{\ifmmode m_{\tau}  \else $m_{\tau}$\fi}}
\newcommand{\dpp}{{\ifmmode \delta_{pert} \else $\delta_{pert}$\fi}}
\newcommand{\dnp}{{\ifmmode\delta_{non-pert}\else$\delta_{non-pert}$\fi}}
\newcommand{\dew}{{\ifmmode \delta_{\rm EW}\else $\delta_{\rm EW}$\fi}}
\newcommand{\rt}{{\ifmmode R_{\tau}  \else $R_{\tau} $\fi}}
\newcommand{\rz}{{\ifmmode R_{Z}  \else $R_{Z} $\fi}}

\newcommand{\swb}{{\ifmmode \sin^2\theta_{\overline{MS}} \else $\sin^2\theta_{\overline{MS}}$\fi}}
\newcommand{\cwb}{{\ifmmode \cos^2\theta_{\overline{MS}} \else $\cos^2\theta_{\overline{MS}}$\fi}}

\begin{document}
\vspace*{4cm}
\title{EGRET Excess of  Galactic Gamma Rays as Signal of Dark Matter Annihilation}

\author{W. de Boer}

\address{Institut f\"ur Experimentelle Kernphysik\\
University of Karlsruhe \\
Postfach 6980 \\
76128 Karlsruhe, Germany\\
E-mail: wim.de.boer@cern.ch}

\maketitle

\abstracts{ The EGRET excess in the diffuse galactic  gamma ray data
above 1 GeV shows all  features expected from Dark Matter WIMP
Annihilation: a)it is present and has the same spectrum in all sky
directions, not just in the galactic plane. b) The intensity of the
excess shows the $1/r^2$ profile expected for a flat rotation curve
outside the galactic disc with additionally   an interesting
substructure in the disc in the form of a doughnut shaped ring at 14
kpc from the centre of the galaxy.  At this radius a ring of stars
indicates the probable infall of a dwarf galaxy, which can explain
the increase in DM density.
From the spectral shape of the excess the WIMP mass is estimated to
be between 50 and 100 GeV, while from the intensity the halo profile
is reconstructed. Given the mass and intensity of the WIMPs the mass
of the  ring can be calculated, which is shown to explain the
peculiar change of slope in the rotation curve at about 11 kpc.
These results are model independent in the sense that only the {\it
known shapes}of signal and background were fitted with free
normalization factors, thus being independent of the model dependent
flux caluclations. The statistical significance is more than
$10\sigma$ in comparison with a fit of the conventional galactic
model to the EGRET data. These signals of Dark Matter Annihilation
are compatible with Supersymmetry including all electroweak
constraints. The statistical significance combined with all features
mentioned above provide an intriguing hint that the EGRET excess is
indeed  a signal from Dark Matter Annihilation.}

\section{Introduction}
Cold Dark Matter (CDM) makes up 23\% of the energy of the universe,
as deduced from the WMAP measurements of the temperature
anisotropies in the Cosmic Microwave Background, in combination with
data on the Hubble expansion and the density fluctuations in the
universe~\cite{wmap}. The Dark Matter has to be much more widely
distributed than the visible matter, since the rotation speeds do
not fall off like $1/\sqrt{r}$, as expected from the visible matter
in the centre, but stay more or less constant as function of
distance. For  a "flat" rotation curve the DM has to fall off slowly
like $1/r^2$ instead of the exponential drop-off for the visible
matter. The fact that the DM is distributed over large distances
implies that its properties must be quite different from the visible
matter, since the latter clumps in the centre owing to its rapid
loss of kinetic energy by the electromagnetic and strong
interactions after infall into the centre.  Since the DM apparently
undergoes little energy loss, it can have at most weak interactions.
In addition its mass is probably large, since it cannot be produced
with present accelerators. Therefore it is generically called a
WIMP, a Weakly Interacting Massive Particle.

Weakly interacting particles can annihilate, yielding predominantly
quark-antiquark pairs in the final state, which hadronize into
mesons and baryons.
 The stable decay and fragmentation
products are neutrinos, photons, protons, antiprotons, electrons and
positrons. From these, the protons and electrons disappear in the
sea of many matter particles in the universe, but the photons and
antimatter particles may be detectable above the background,
generated by  particle interactions. Such  searches for indirect
Dark Matter detection have been actively pursued, see e.g the review
by Bergstr$\rm \ddot{o}$m \cite{bergstrom} or more recently by
Bertone, Hooper and Silk.\cite{Bertone:2004pz}

 The present analysis on diffuse galactic gamma rays differs from previous ones
by considering simultaneously the complete sky map {\it and} the
energy spectrum, which allows us to constrain both the halo
distribution {\it and} the WIMP mass. More details have been given
elsewhere \cite{deboer1,deboer2,deboer3}.
The constraint on the WIMP annihilation cross section from WMAP is
discussed in Section 2, while the constraints on the mass and the DM
halo profile from the EGRET excess are discussed in Sections 3. The
 summary is given in Section 4.

\section{Annihilation Cross section Constraints from WMAP}
In the early universe all particles were produced abundantly and
were in thermal equilibrium through annihilation and production
processes. At temperatures below the mass of the WIMPS the number
density drops exponentially. The annihilation rate $\Gamma=<\sigma
v> n_\chi$ drops exponentially as well, and if it drops below the
expansion rate, the WIMP's cease to annihilate. They fall out of
equilibrium (freeze-out) at a temperature of about $m_\chi/22$
 \cite{kolb} and a relic cosmic abundance remains.

For the case that $<\sigma v>$ is energy independent, which is a
good approximation in case there is no coannihilation, the present
mass density in units of the critical density is given by
\cite{jungman}: \bq \Omega_\chi h^2=\frac{m_\chi
n_\chi}{\rho_c}\approx (\frac{2\cdot 10^{-27} cm^3 s^{-1}}{<\sigma
v>})\label{wmap}.\eq One observes that the present relic density is
inversely proportional to the annihilation cross section at the time
of freeze out, a result independent of the WIMP   mass (except for
logarithmic corrections). For the present value of $\Omega_\chi
h^2=0.113\pm0.009$ the thermally averaged total cross section at the
freeze-out temperature of $m_\chi/22$ must have been around $2\cdot
10^{-26} {\rm cm^3s^{-1}}$. The observed annihilation rate will be
compared with  this generic cross section, which basically only
depends on the expansion rate of the universe, i.e. on the value of
the Hubble constant. However, it should be noted that this cross
section may be energy dependent and the annihilation cross section
in the present universe may be much smaller than the value deduced
from the time of freeze out, when the temperature was
$m_\chi/22\approx $ several GeV. On the other hand the annihilation
rate may be enhanced by the clustering of DM in ``microhaloes'',
which increases the density locally. This unknown enhancement
factor, usually called ''boost factor'', may vary from  a few to a
few thousand.\cite{dokuchaev,moore}

\section{Indirect Dark Matter Detection}\label{sec2}
The neutral particles play  a very special role for indirect DM
searches, since they point back to the source.  The charged
particles change their direction by the interstellar magnetic
fields, energy losses and scattering. Therefore the gamma rays
provide a perfect means to reconstruct the intensity (halo) profile
of the DM by observing the intensity of the gamma ray emissions in
the various sky directions. Of course, this assumes that one can
distinguish the gamma rays from DM annihilation  from the
background, mainly from proton-proton interactions. Both for DMA and
pp collisions the gamma rays originate mainly from the decay of
neutral pions, a light particle produced abundantly in the
hadronization process of quarks into hadrons. However, the protons
in the galaxies and consequently the quarks inside the protons have
a steeply falling energy spectrum ($N\propto E^{-2.7}$). In
contrast, the quarks from DM annihilation are mono-energetic, since
the WIMPS  annihilate almost at rest, so their mass is converted
completely into kinetic energy of the much lighter quarks. Each
quark thus obtains an energy corresponding to the mass of the WIMP,
which yields a gamma ray spectrum with a sharp cut-off at the mass
of the WIMP. So from the shape of the spectrum the WIMP mass can be
deduced.  The difference in spectral shape between DMA and
background allows to obtain their absolute normalizations by fitting
their shapes to the EGRET data. These shapes are well known from
accelerator experiments and can be obtained e.g. from the  PYTHIA
code for quark fragmentation \cite{pythia}; the parameters in this
code have been optimized to fit a wide variety of accelerator data
with a single model, the string fragmentation model.
 The fit of the normalizations can
be repeated in many different sky direction to obtain the halo
profile of the DM. Given the WIMP {\it number density} in all
directions from the flux of the excess and the WIMP {\it mass} from
the spectrum allows to reconstruct the DM mass distribution in our
galaxy, which in turn can be used to reconstruct the rotation curve.

A very detailed gamma ray distribution over the whole sky was
obtained by the Energetic Gamma Ray Emission Telescope EGRET, one of
the four instruments on the Compton Gamma Ray Observatory CGRO,
which collected data during nine  years, from  1991 to 2000.  The
EGRET telescope was carefully calibrated in the energy range of 0.1
to 30 GeV, but using Monte Carlo  simulations the energy range was
recently extended up to 120 GeV \cite{optimized}  with a
correspondingly larger uncertainty, mainly from the self-vetoing of
the detector by the back-scattering from the electromagnetic
calorimeter into the veto counters for high energetic showers. It
was already noticed in 1997 that the EGRET data showed an excess of
gamma ray fluxes for energies above 1 GeV if compared with
conventional galactic models.\cite{hunter}

\begin{figure}
\begin{center}
 \includegraphics [width=0.32\textwidth,clip]{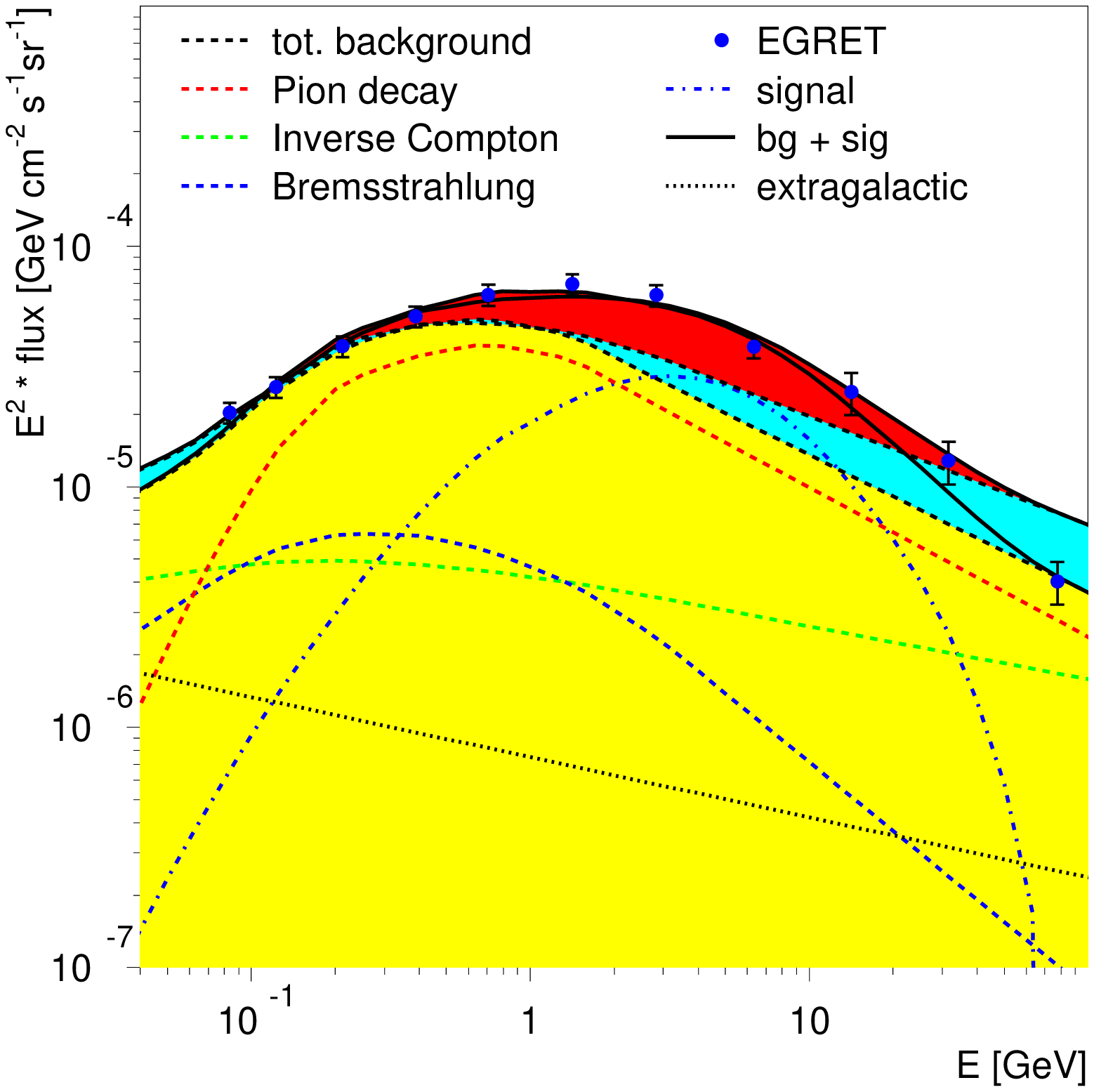}
 \includegraphics [width=0.32\textwidth,clip]{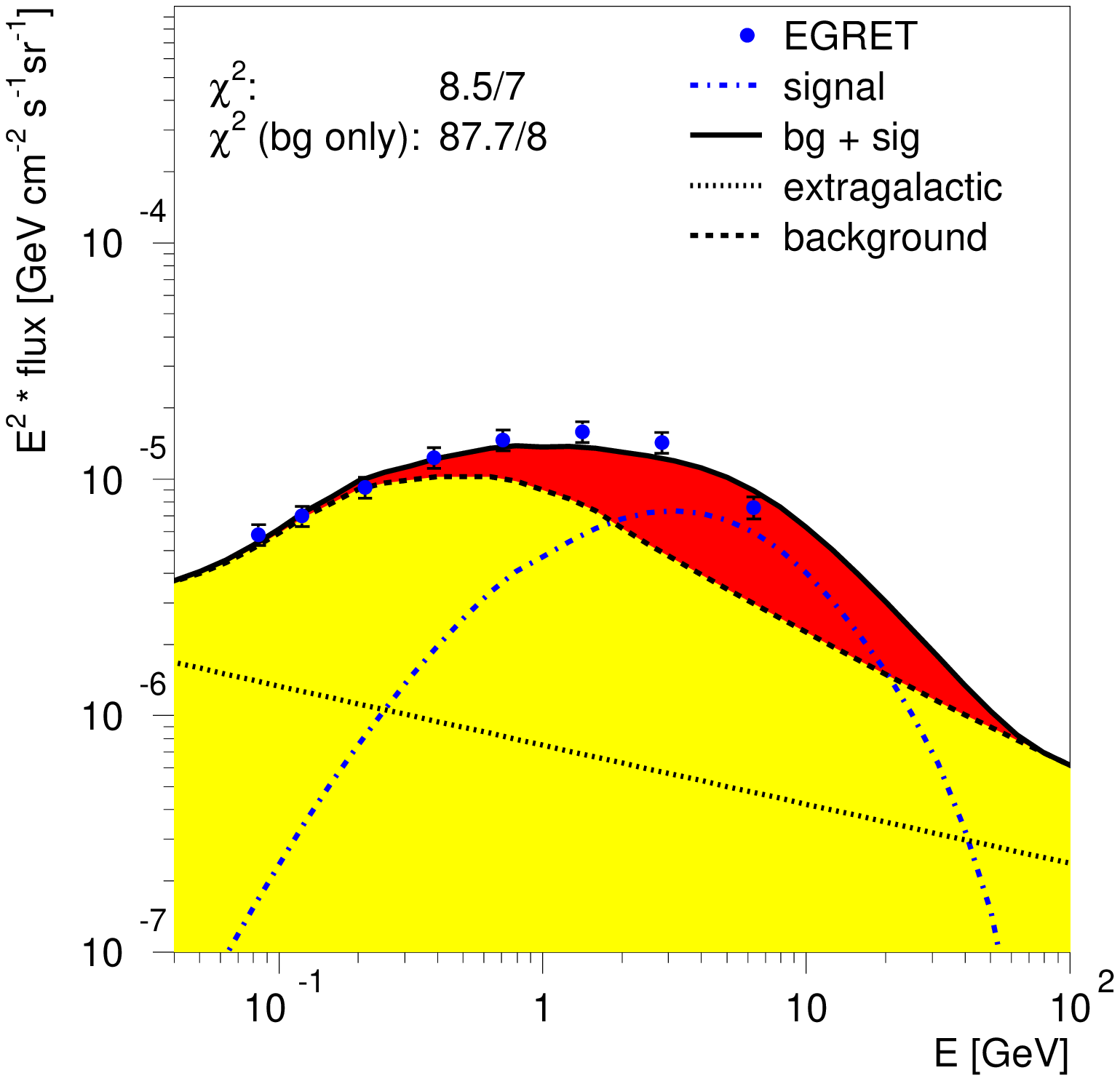}
 \includegraphics [width=0.32\textwidth,clip]{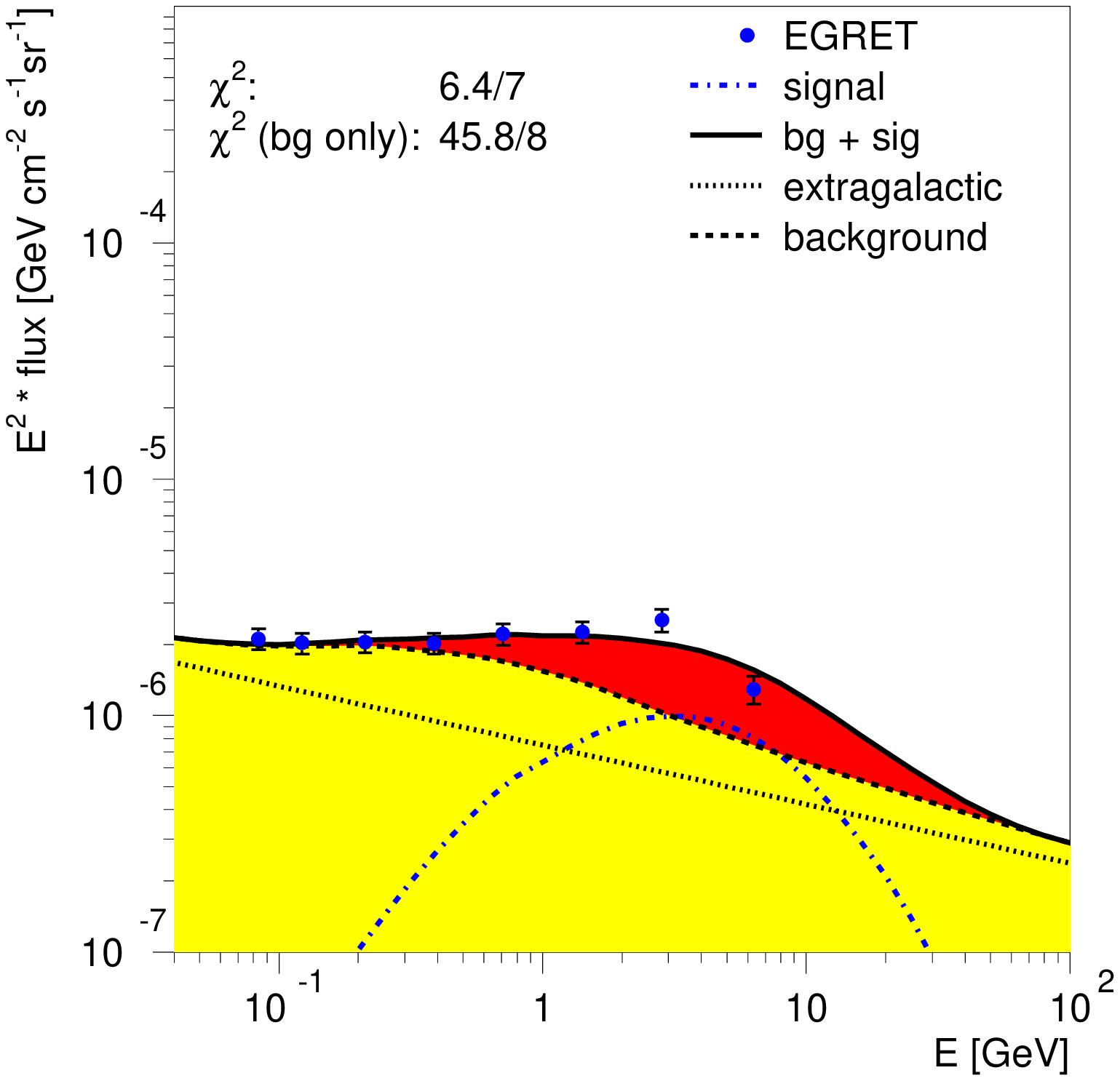}
 \caption[]{
 The diffuse gamma-ray energy spectrum of 3 angular regions: from left to right:
  towards the galactic centre (latitudes  $0^\circ<|b|<5^\circ$; longitudes $0^\circ<|l|<30^\circ$),
   the galactic anticentre ($0^\circ<|b|<10^\circ$;  $90^\circ<|l|<270^\circ$)and the pole regions
   ($60^\circ<|b|<90^\circ$;  $0^\circ<|l|<360^\circ$),
  as measured by the EGRET space telescope.
In the two panels on the right the solid straight line represents
the fitted contribution from the extragalactic background, while the
dotted line indicates the contribution from the annihilation from 65
GeV WIMPs. The  total background (DMA)  is
   indicated by the light (yellow)   (dark (red))  shaded area, respectively.
   In the panel on the left the various contributions to the
background are indicated as well, while
  the  uncertainties
  from the background  are indicated  by the medium shaded (blue) area.
  Here the upper  edge of the
medium shaded (blue) area corresponds the hardest  spectrum
 from Kamae et al.\cite{kamae} with the power index of 2.5, while
the lower edge corresponds to the shape of the conventional GALPROP
model.\cite{optimized} Note that since the background normalization
is left free, the low energy data (where only the background
contributes) are always well fitted and different shapes only show
up at larger energies.
 \label{excess}}
\end{center}
\end{figure}

 Fitting the three contributions of galactic background, extragalactic background and DMA to
 the energy spectra of 180 independent sky directions yielded
 astonishingly good fits with the free normalization of the background agreeing reasonably
 well with the absolute predictions  of the galactic models \cite{galprop,galprop1}
 for the energies between 0.1 and 0.5 GeV. Above these energies a clear contribution from Dark
Matter annihilation is needed,  but the excess in different sky
directions can be explained by a single WIMP mass. The fits
  for 3 different sky directions are shown in Fig. \ref{excess}.

 Alternative explanations for the excess have been plentiful. Among
them: weak point sources, which could not be resolved from the
background by the EGRET satellite. This is unlikely, since the point
sources usually have a rather soft spectrum. If one assumes that
most of the unresolved point sources would have   similar spectra,
their subtraction would reduce the observed diffuse spectra below 1
GeV,  but the data above 1 GeV would be much less affected. With our
fitting procedure of the shapes, the background is determined by the
data below 1 GeV and would thus become lower with unresolved point
sources subtracted. Thus would lead to an even stronger excess!
\begin{figure}
\begin{center}
 \includegraphics [width=0.33\textwidth,clip]{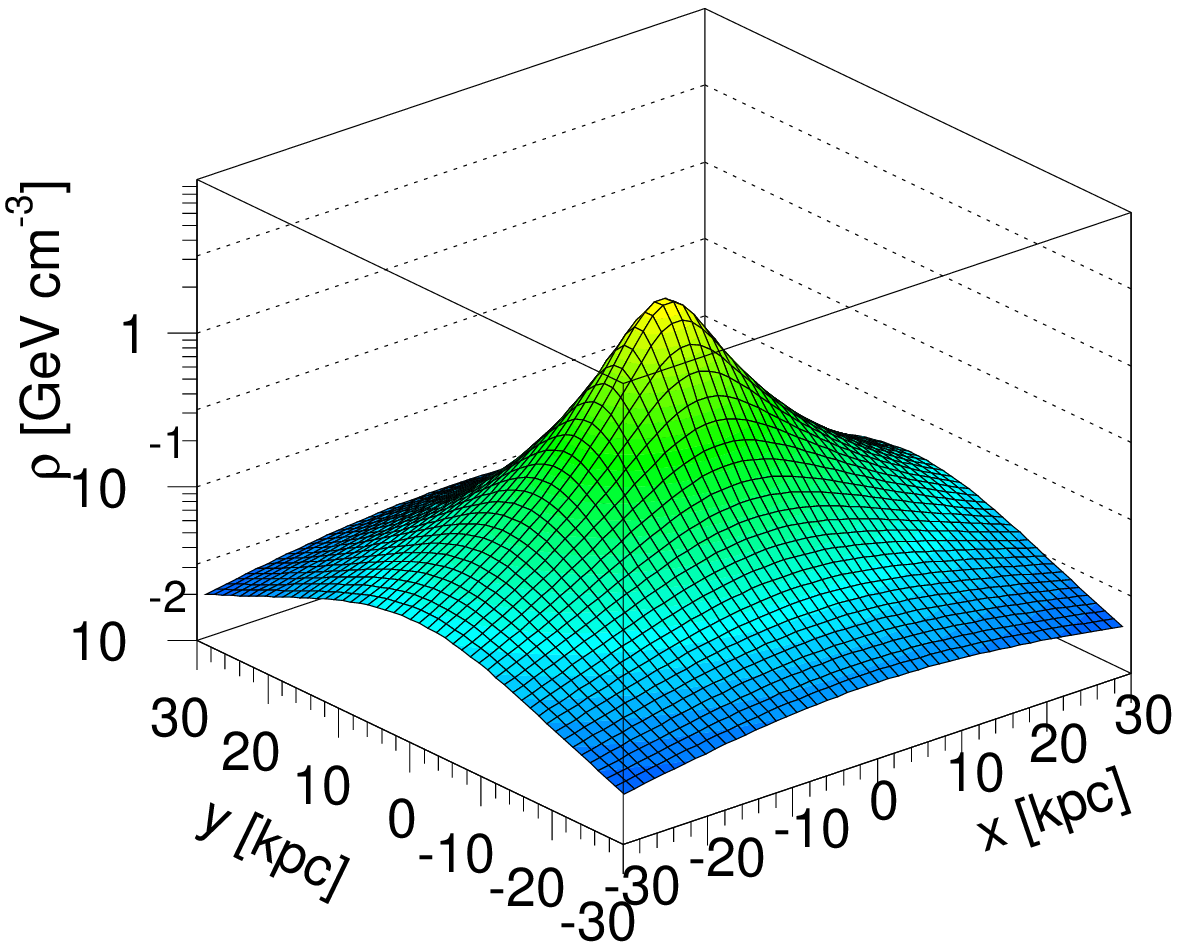}
\includegraphics [width=0.33\textwidth,clip]{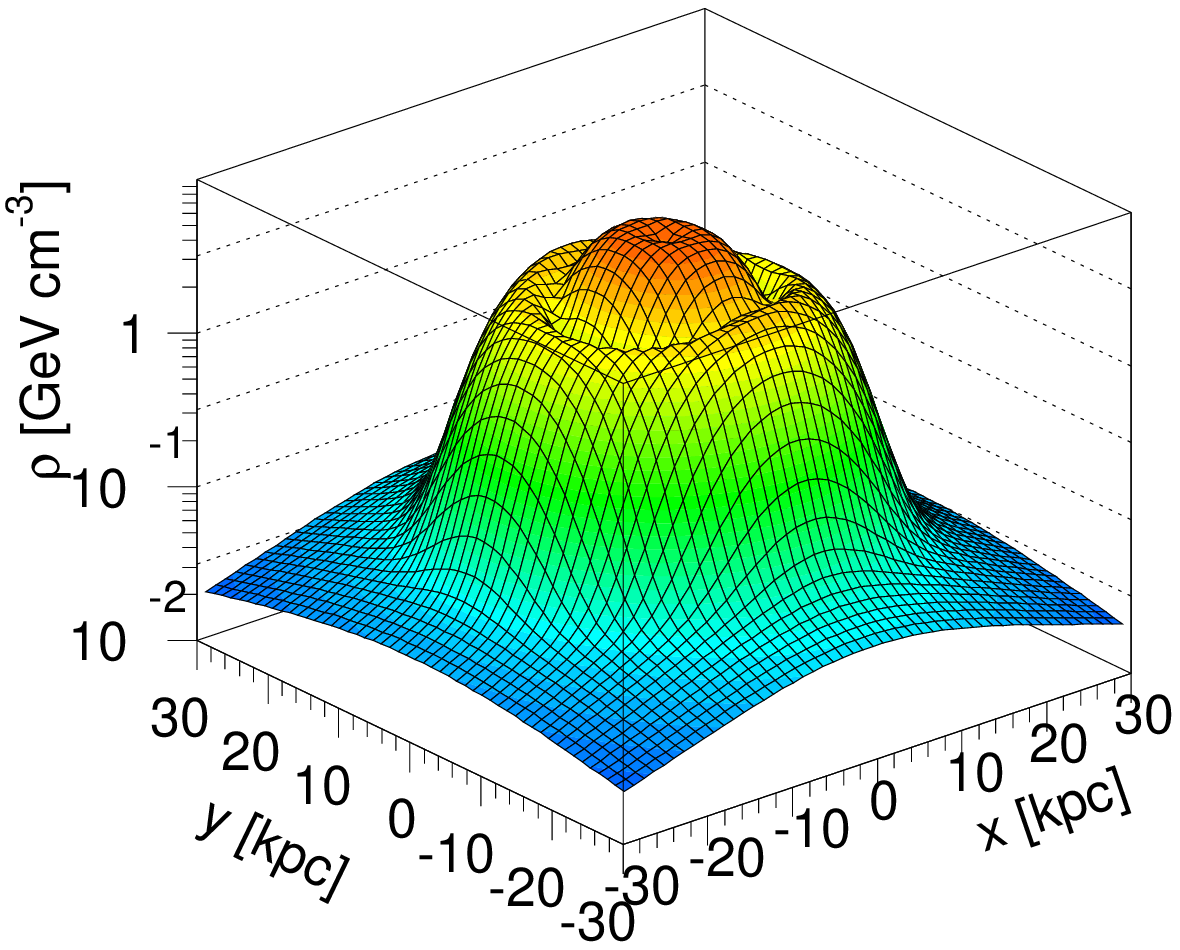}
\includegraphics [width=0.33\textwidth,clip]{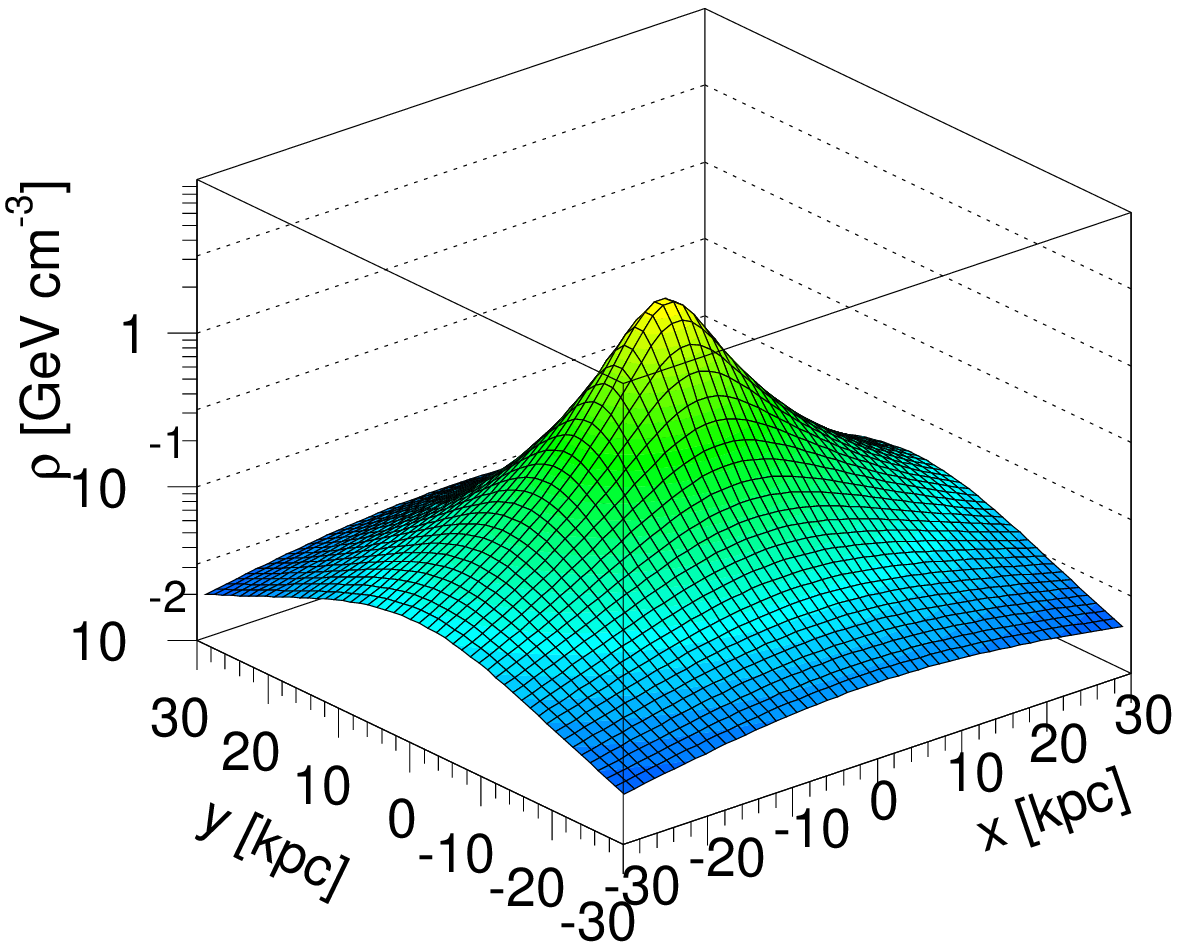}
\includegraphics [width=0.33\textwidth,clip]{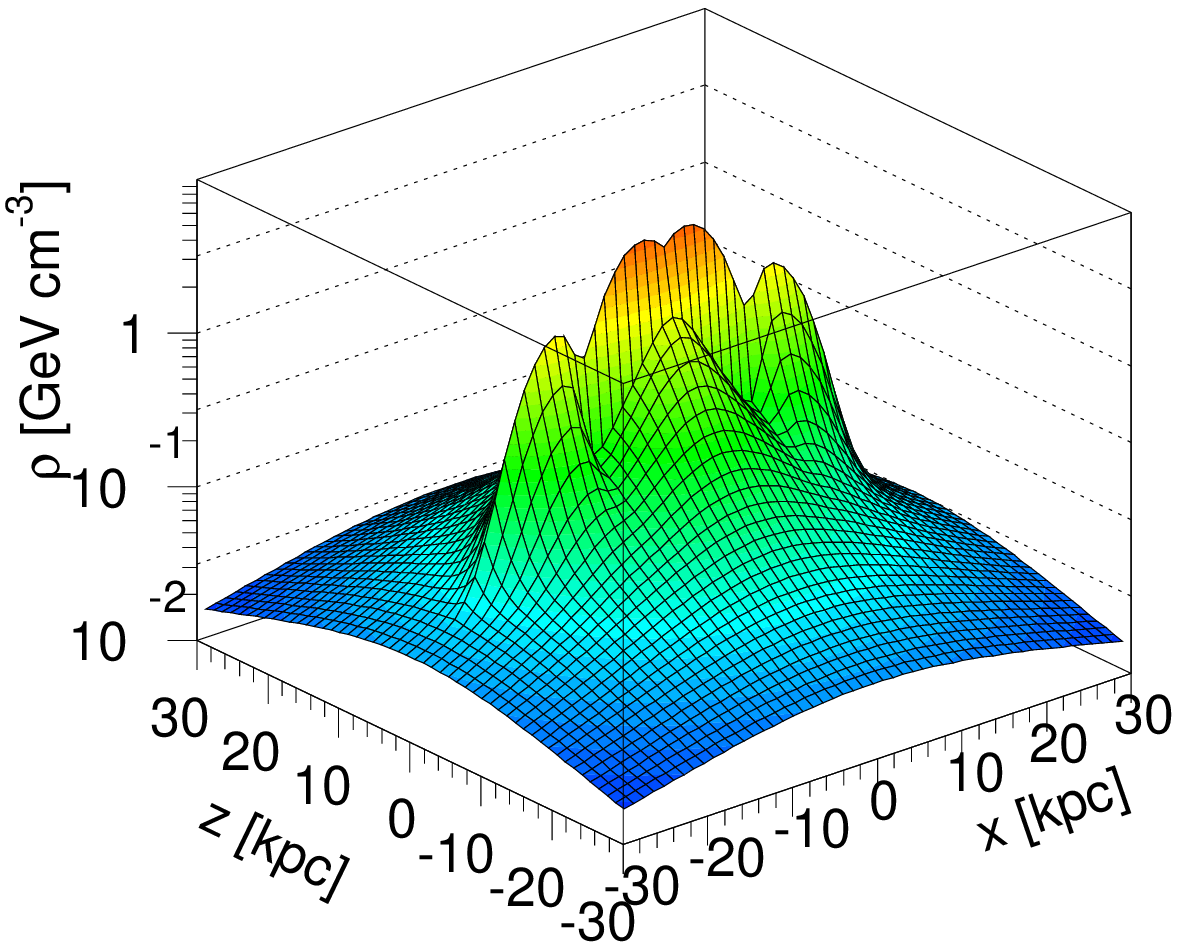}
 \caption[]{
  3D-distributions of the $1/r^2$ haloprofile
  in the galactic xy-plane  (top row) and xz-plane (bottom row)
  without (left) and with (right) rings.
}
 \label{profile}
\end{center}
\end{figure}

 Other ways to increase the excess would be to harden the spectra of
 the primary nuclei and electrons with respect to the locally
 measured spectra. Inhomogeneities in the spectra could happen e.g.
 by density fluctuations from the spiral arms or Supernovae
 explosions.
 A  summary of these discussions has been given by Strong et
al..\cite{optimized} They find that by modifying the electron and
proton  spectra simultaneously, they can improve the description of
the data. However,  above 2 GeV the predicted flux of this so-called
``optimized'' model is still too low, as shown in Fig. 9 of their
paper. Since they tried to predict the absolute flux, the overall
normalization errors are plotted. However, if one only considers the
shape of the spectra, then only the relative systematic errors
between the energy points play a role and these are at least a
factor two smaller.  In this case the probability of the fit, if the
shape of the optimized model is fitted to all sky directions, is
below $10^{-7}$.\cite{sander} Adding DM to the optimized model
improves the fit probability  to 0.8 \cite{sander}, of course with a
lower boost factor (about factor three), but still a need for DM is
evident.
 Similar results are obtained for the
shape proposed by Kamae et al..\cite{kamae} Here the reduction of
the boost factor is considerably less, mainly because these authors
try to improve the fit by changing the proton spectra only, while in
the optimized model both the electron spectra and proton spectra are
modified.

An alternative way of formulating the problems of the models without
DMA: if the shape of the EGRET excess can be explained perfectly in
all sky directions by a gamma contribution originating from the
fragmentation of mono-energetic quarks, it is very difficult to
replace such a contribution by an excess from nuclei (quarks) (or
electrons) with a steeply falling energy spectrum.
\begin{figure}
\begin{center}
 \includegraphics [width=0.45\textwidth,clip]{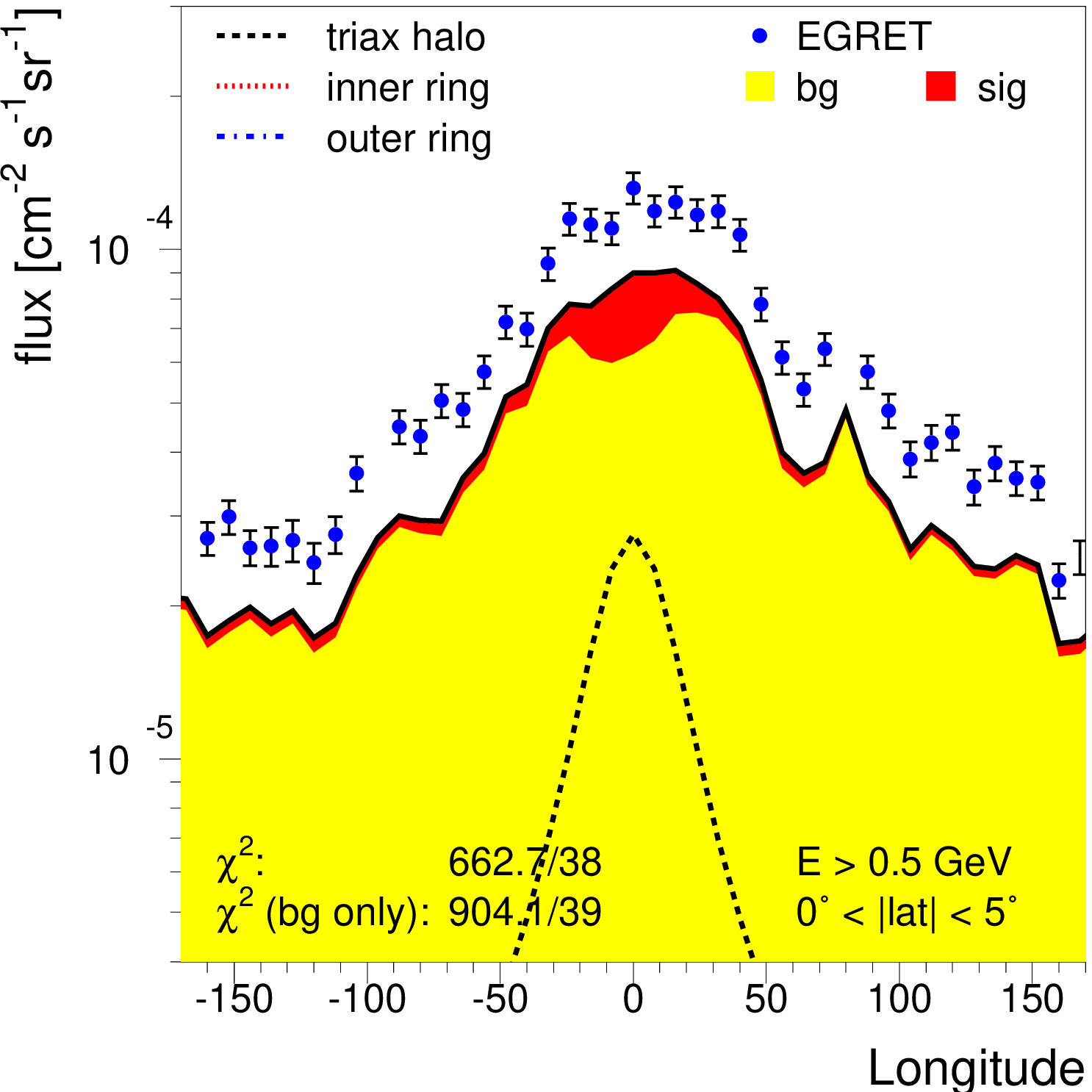}
 \includegraphics [width=0.45\textwidth,clip]{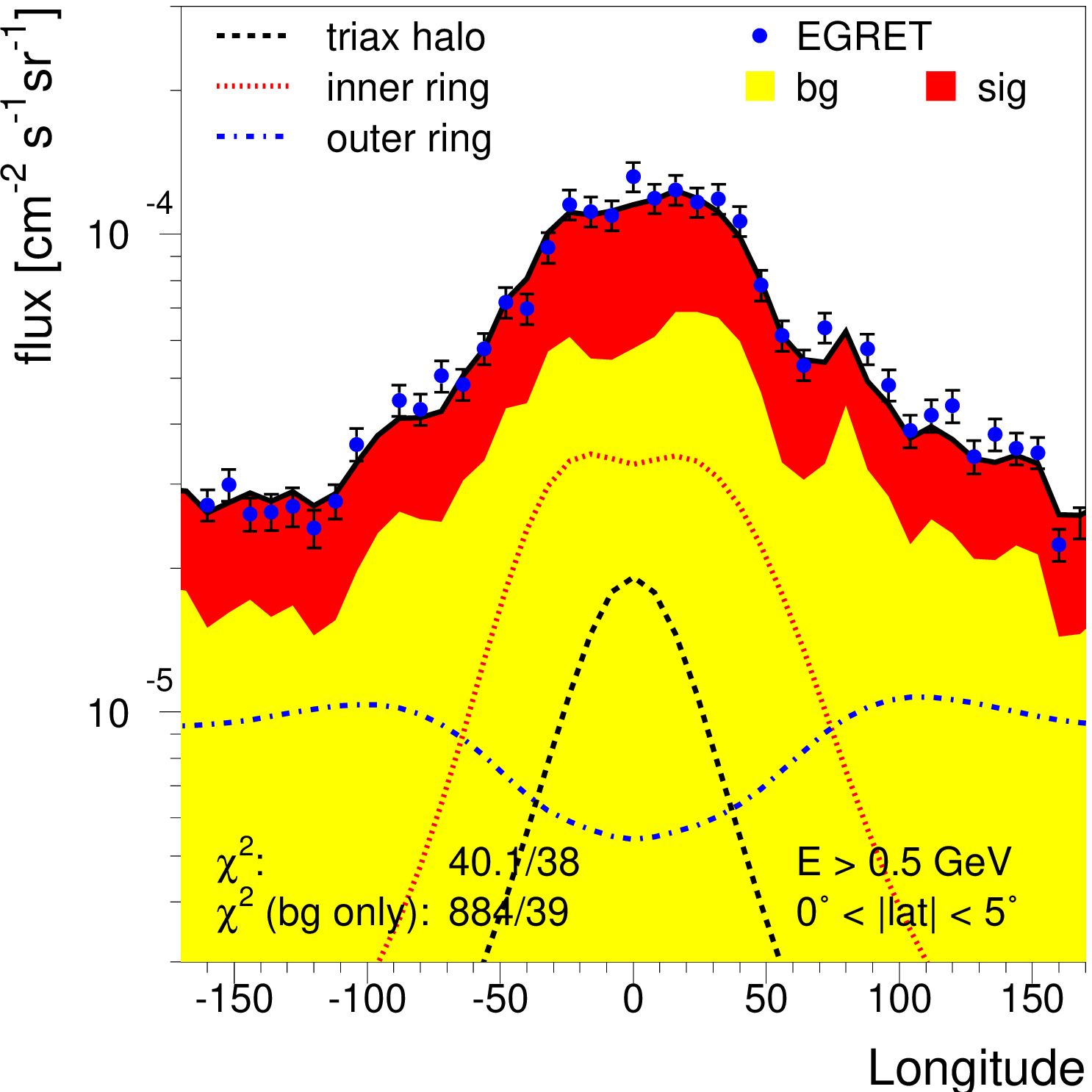}
 \includegraphics [width=0.45\textwidth,clip]{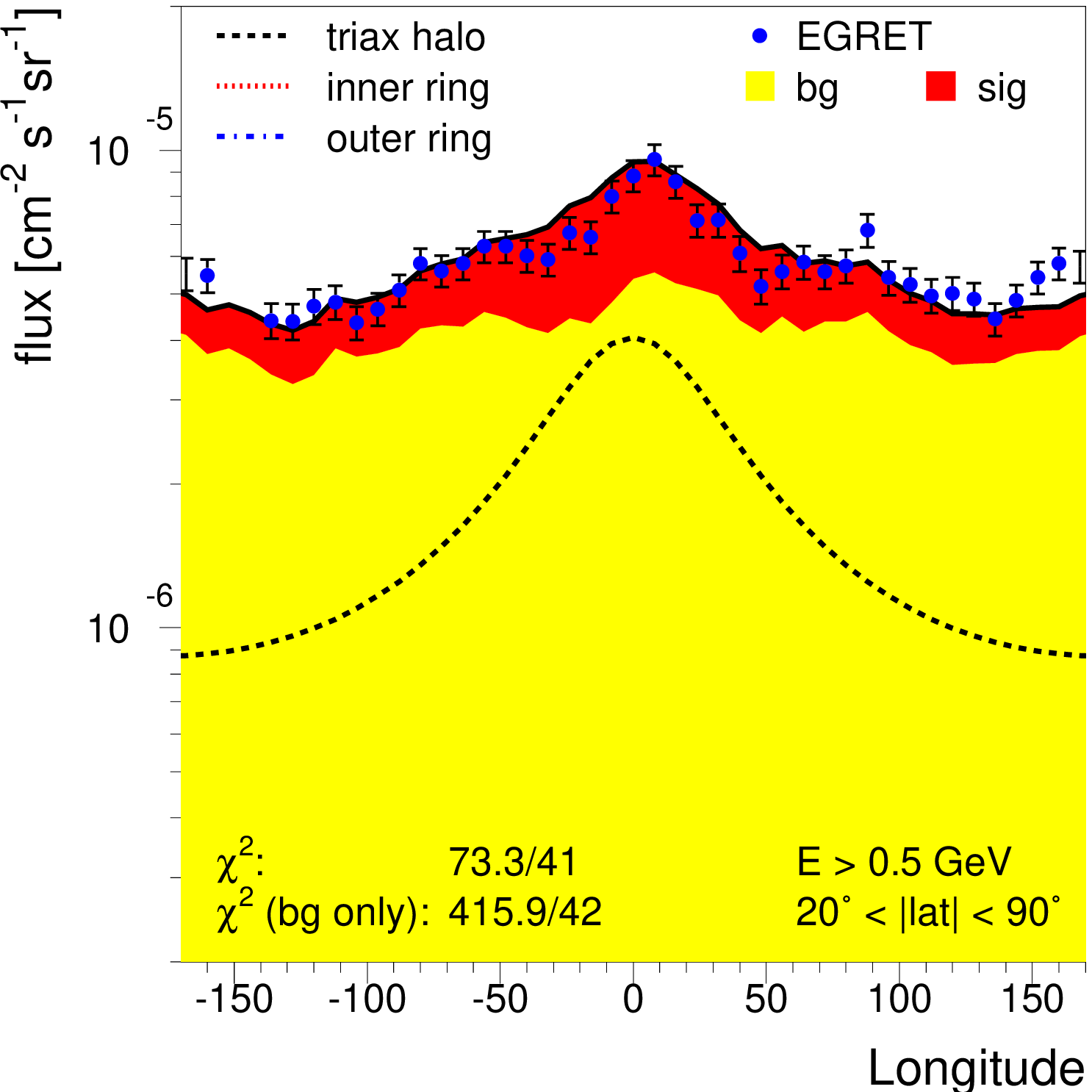}
 \includegraphics [width=0.45\textwidth,clip]{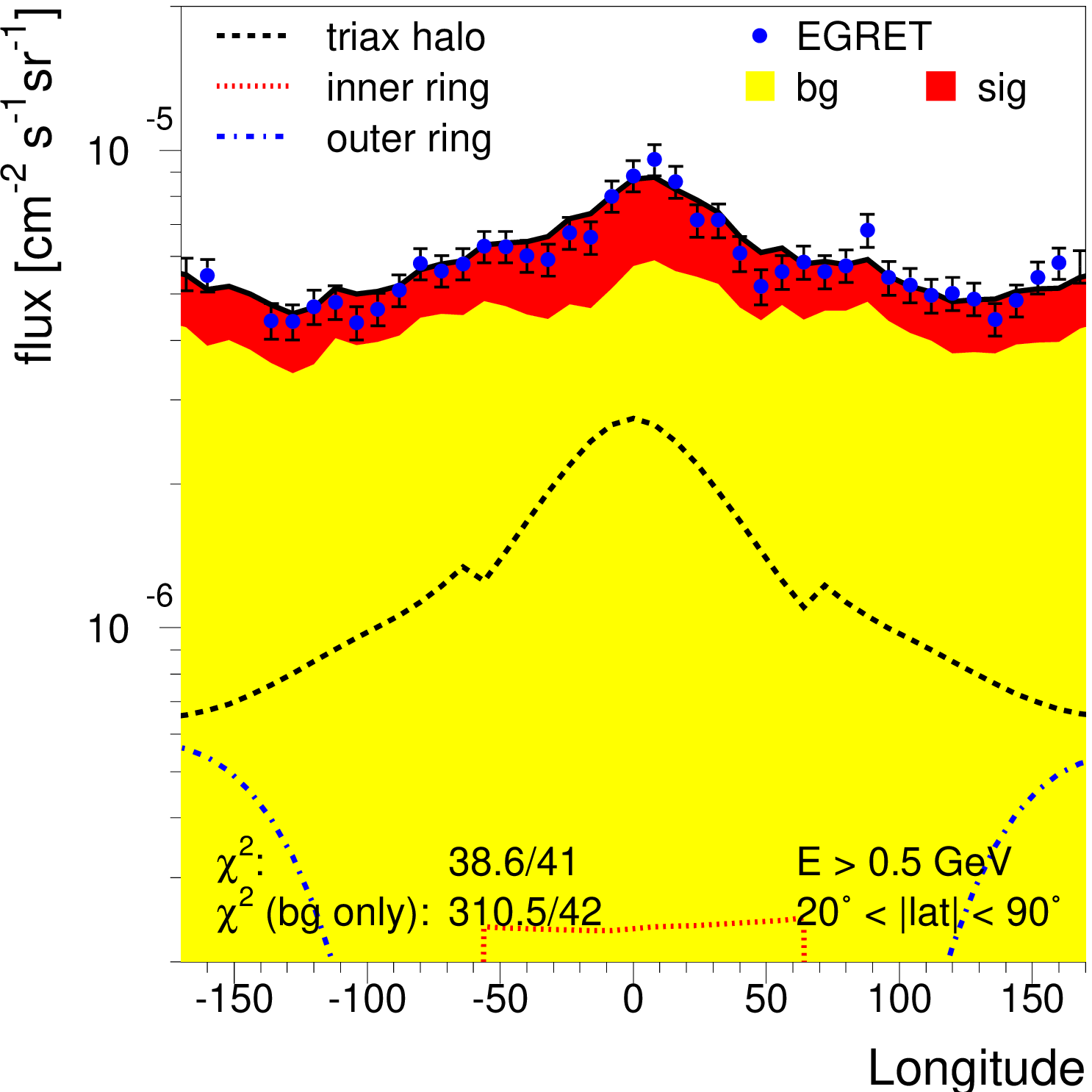}
 \caption[]{Top row:
 the longitude distribution of diffuse gamma-rays in the disc of the galaxy
 (latitudes $0^\circ<|b|<5^\circ$)
for the $1/r^2$ profile without (left) and with rings (right).
 The points represent the EGRET data.
  Bottow row: as above for the polar regions of our galaxy (latitudes  $20^\circ<|b|<90^\circ$) .}
 \label{long}
\end{center}
\end{figure}

From the excess in the various sky directions one can obtain the
halo profile under the assumption that the clustering of the DM is
similar in all sky directions. This is not necessarily true, since
near the centre of the galaxy   clumps may be tidally disrupted by
the flyby of stars. The annihilation rate is in general proportional
to $B\rho^n$, where B is the boost factor and $n$ is between 1 and
2, depending on how much of the DM is clustered (n=2 for no
clustering and n=1 if all DM is in clusters). Since the EGRET excess
measures only the product $B\rho^n$, several choices can be made.
For definiteness we use $n=2$ and $B$ to be the same for all
directions and an isothermal halo, which falls like $1/r^2$ as
expected for a flat rotation curve.  The result is surprising: in
addition to the isothermal profile the EGRET excess show a
substructure in the form of toroidal rings at 4 and 14 kpc, as shown
in Fig. \ref{profile}: on the left hand side the contribution from
the $1/r^2$ profile is shown, while for the right hand side the ring
structure is added. Such enhanced gamma radiation at 4 and 14 kpc
was already observed in the original paper on the EGRET
excess.\cite{hunter} Note that the appearance of substructure would
also be obtained if a radial dependence of  $n$ and $B$ would have
been taken. The analysis is sensitive to the radii of ringlike
structures, since we are not located at the centre: assuming a
constant flux along the ring yields automatically more flux from the
nearest parts. The need for these additional rings is most easily
seen by comparing the longitudinal profiles in the galactic plane
and towards the galactic poles. As shown in Fig. \ref{long} the pole
regions are described reasonably well without rings, but for the
galactic plane the $1/r^2$ profile only describes the data towards
the centre. For the larger latitudes one needs the rings, as
indicated by the right top panel. Note that for each bin only the
flux integrated for data above 0.5 GeV has been plotted.

The position and shape of the outer ring coincides with the ring of
stars, discovered in 2003 by several
groups.\cite{yanny,ibata,rocha-pinto1}  These stars show a much
smaller velocity dispersion (10-30 km/s) and larger z-distribution
than the thick disc, so it cannot be considered an extension of the
disc. A viable alternative is the infall of a dwarf galaxy
 \cite{yanny,martinez-delgado}, for which one expects in addition to
he visible stars a DM component. From the size of the ring and its
peak density one can estimate the amount of DM in the outer ring to
be  $\approx 10^{10}-10^{11}$ solar masses. Since the gamma ray
excess requires the full 360$^\circ$ of the sky, one can extrapolate
the observed $100^\circ$ of visible stars to obtain
 a total mass of  $\approx 10^8-10^9$ solar masses \cite{yanny,ibata}, so the
baryonic matter in the outer ring is only a small fraction of its
total mass.
\begin{figure}
\begin{center}
 \includegraphics [width=0.5\textwidth,clip]{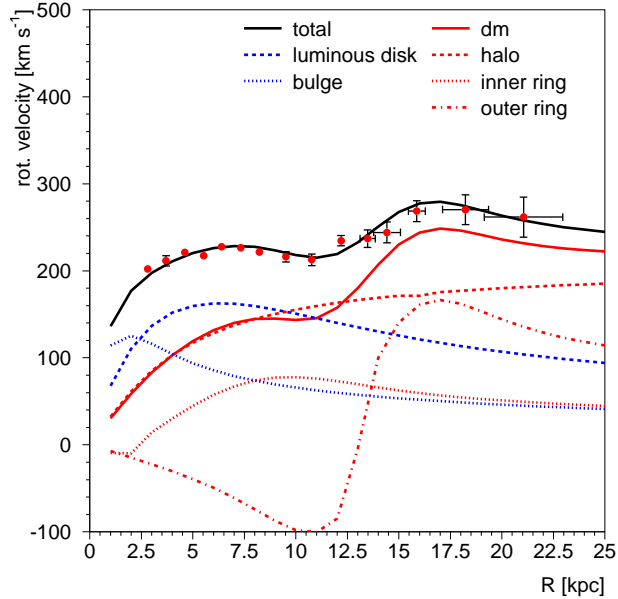}
 \caption[]{
  The rotation curve from our galaxy with the DM contribution determined from the
   EGRET excess of diffuse gamma rays. The data are averaged from Ref. \cite{deboer3}.}
 \label{rot}
\end{center}
\end{figure}

The inner ring at 4.2 kpc with a width of 2.1 kpc in radius and  0.2
kpc in $z$ is more difficult to interpret, since the density of the
inner region is modified by adiabatic compression
 and  interactions between
the bar and the halo.
However, it is interesting to note that its coordinates coincide
with the ring of cold dense molecular hydrogen gas, which reaches a
maximum density at 4 kpc and has a width of 2 kpc as
well.\cite{hunter} Molecules form from atomic hydrogen in the
presence of dust or heavy nuclei. So a ring of neutral hydrogen
suggests an attractive gravitational potential in this region, in
agreement with the EGRET excess.

To prove that the enhanced gamma ray density is indeed connected to
non-baryonic mass the rotation curve was reconstructed from the
excess of the diffuse gamma rays in the following way: since the
flux determines the number density of DM for a given boost factor
and since the mass of each WIMP is between 50 and 100 GeV, one can
determine the relative masses of the components (rings plus
spherical part)  and consequently predict the shape of the rotation
curve. The absolute value of the mass can be obtained by requiring
that the rotation speed of the solar system is 220 km/s at 8.5 kpc.
The two ring model describes the peculiar change of slope at 11 kpc
well, as shown in Fig. \ref{rot}. The contributions from each of the
mass terms have been
 shown separately. The basic explanation for the negative contribution from the outer ring
 is that a tracer star at the
 inside of the ring at 14 kpc feels an outward force from the ring, thus a negative
 contribution to the rotation velocity.
 It has often been argued that  the outer
 rotation curve cannot be taken seriously, because
the errors are large due to the fact that the absolute values of the
rotation velocities strongly depend on the value of $R_0$, the
distance between the solar system and the galactic centre. This is
true, as shown by Honma and Sofue\cite{honma}, but they show that
the {\it change in slope} at about 1.3$R_0$  is independent of
$R_0$. In addition, it has been argued that the inner and outer
rotation curve are difficult to compare, since the methods are
completely different. The methods are indeed different, but
 the first 3 data points from the outer rotation curve
  (between 8 and 11 kpc) show the same slope as the ones from the
  inner rotation curve, so there seems to be no systematic effect
  related to the different methods.

\section{Summary and Outlook}
In summary, the EGRET data shows an intriguing hint of DM
annihilation, since it explains many unrelated facts simultaneously:

a) An excess of diffuse galactic gamma rays which shows a {\it
spectrum} consistent with the expectation from WIMP annihilation
into gamma rays originating from the fragmentation of
{mono-energetic} quarks.

b) The excess is present in {\it all} sky directions with the same
spectrum, thus excluding that it originates from anomalous
contributions in the centre of the galaxy.

c) The excess shows an strongly increased intensity at positions
where extra DM is expected, namely at two doughnut shaped structures
at radii of 14 and 4 kpc from the centre of the galaxy. At 14 kpc
one has observed a ring of stars thought to originate from the
infall of a dwarf galaxy, while at 4 kpc one finds an enhanced
concentration of molecular hydrogen thought to form from atomic
hydrogen in the presence of dust or heavy nuclei, which can be
collected in the gravitational potential of a ring of DM.

d) The enhanced excess of gamma rays cannot be due to additional gas
in these rings as proven by the rotation curve calculated from the
gamma ray excess: the mass in the rings perfectly describe the
hitherto unexplained change of slope in the rotation curve at a
distance of about 11 kpc. The amount of  visible matter is far too
low to have such an impact on the rotation curve.
\begin{figure}
\begin{center}
 \includegraphics [width=0.45\textwidth,clip]{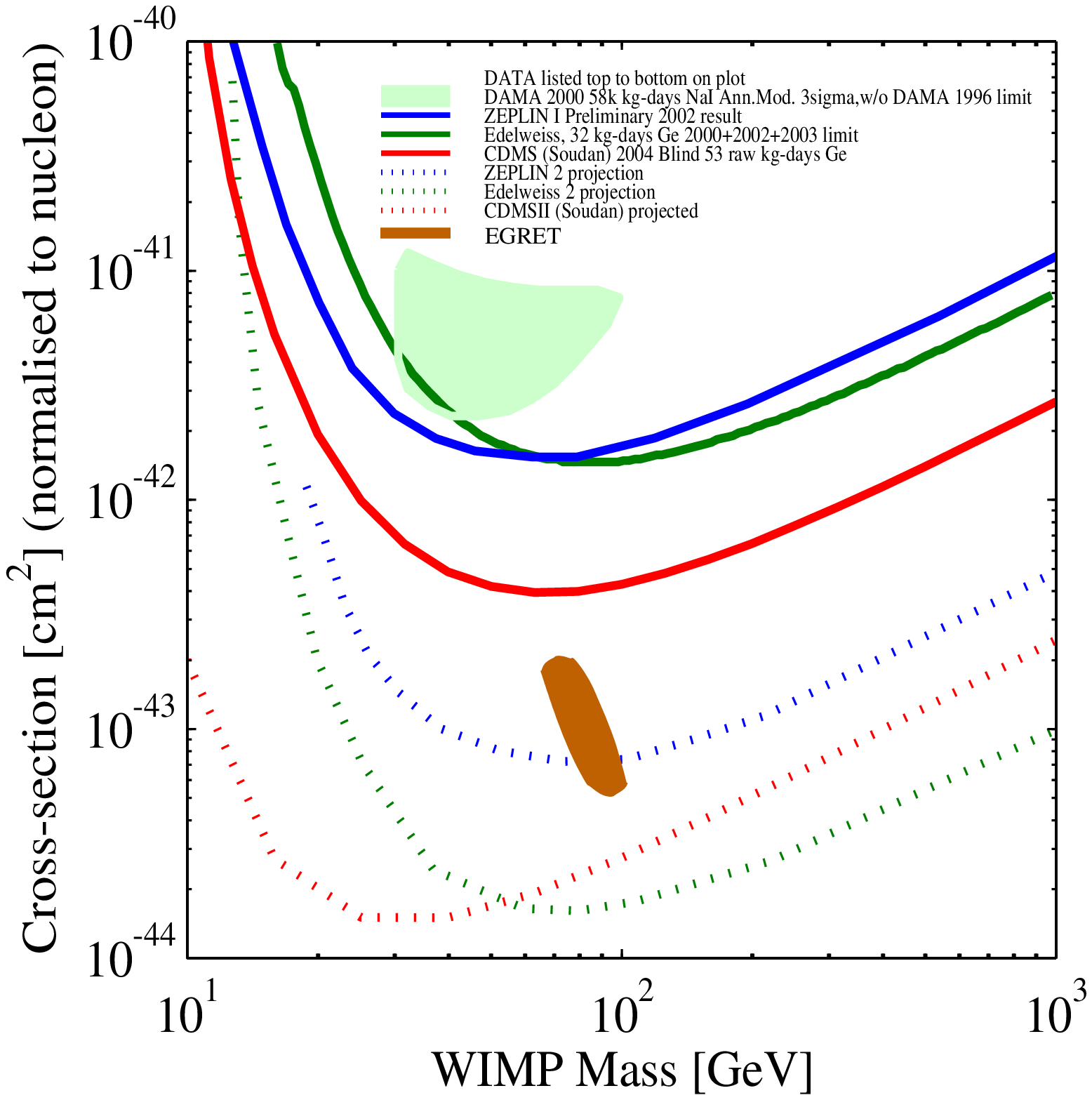}
\includegraphics [width=0.45\textwidth,clip]{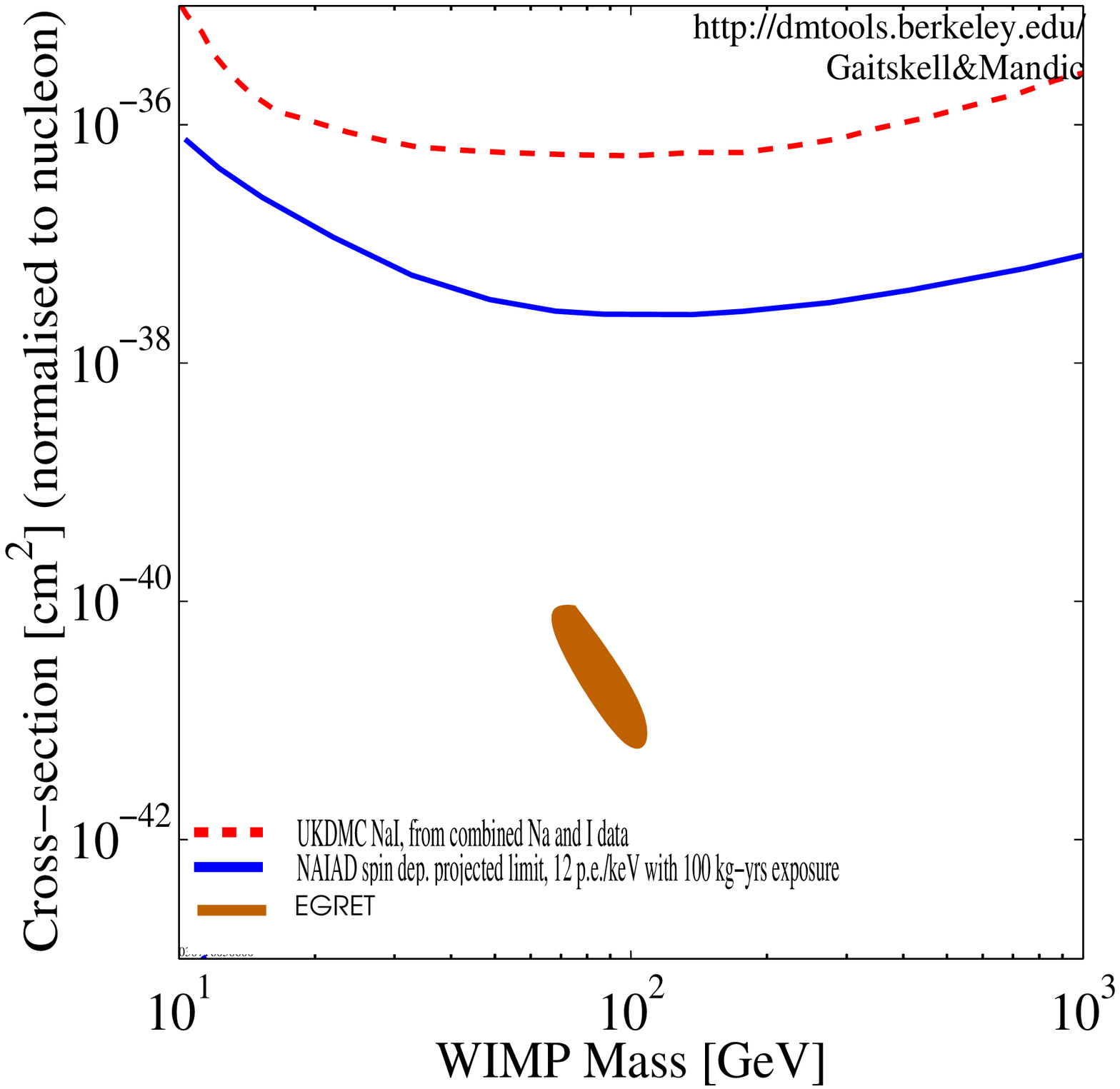}
 \caption[]{
 The spin-independent (left) and spin-dependent (right)
 neutralino-nucleon cross section  as function of the neutralino
 mass for the SUSY parameters from this analysis\cite{deboer3} (oval
 shaded (brown) area in comparison with results from present and future direct DM
 detection experiments.
 }
 \label{direct}
\end{center}
\end{figure}

The results mentioned above make no assumption on the nature of the
Dark Matter, except that its annihilation produces hard gamma rays
consistent with the fragmentation of monoenergetic quarks between 50
and 100 GeV. WIMPs produce such monoenergetic quarks with energies
equal to the WIMP mass. WIMP masses in this range and the observed
WIMP self annihilation cross section  are consistent with the
Lightest Supersymmetric Particle predicted in the Minimal
Supersymmetric Model with supergravity inspired symmetry breaking,
called the mSUGRA model, if one assumes the enhancement of the
annihilation by the clustering of DM to be of the order of 50, which
is an order of magnitude not unexpected.\cite{dokuchaev,moore}

Within this supersymmetric model one finds a spin-independent cross
section for elastic scattering of a WIMP on a proton of about
$10^{-43}~\rm cm^2$, which is within reach\cite{gait} of future
experiments as shown in Fig. \ref{direct}. This elastic scattering
cross section was calculated with Darksusy\cite{darksusy}.

Direct and indirect detection experiments do not prove the
supersymmetric nature of the WIMPs. If the WIMPs are indeed the
lightest supersymmetric particle, then this will become clear at the
future LHC collider under construction at CERN in Geneva, where
supersymmetric particles of the mass range deduced from the EGRET
data\cite{deboer3} should be observable from 2008 onwards, if they
exist.

In our analysis we only fit the known spectral shapes of the various
processes with arbitrary normalizations, so  the analysis becomes
largely model independent. Interestingly,  the normalization factors
come out to be in  agreement with expectations, both for the WIMP
signal and the background.

Alternative models for the EGRET excess without DM  have to assume
that the locally measured fluxes of protons and electrons are not
representative for our galaxy. These models provide significantly
worse fits to the data, if one takes the strong correlations in the
errors between the different energy bins into account. Of course
such models do not explain  the stability of the ring of stars at 14
kpc and the change of slope in the rotation curve at $r=1.3R_0$.

Therefore the statistical significance of the EGRET excess of at
least 10 $\sigma$, if fitted to the shape of the diffuse gamma ray
background only, combined with  all   features mentioned above
provides an intriguing hint that this excess is indeed  indirect
evidence for Dark Matter annihilation.
\section{Acknowledgements}
I thank my close collaborators A. Gladyshev, D. Kazakov, C. Sander
and V. Zhukov for their contributions to this interesting project.
Furthermore  I thank V. Moskalenko, A. Strong and  O. Reimer  for
numerous discussions on galactic gamma rays and analysis of EGRET
data. This work was supported by the DLR (Deutsches Zentrum f\"ur
Luft- und Raumfahrt)
 and a grant from the
DFG (Deutsche Forschungsgemeinschaft, Grant 436 RUS 113/626/0-1).

\end{document}